\documentstyle[aps,
prl]{revtex}

\newcommand{\fe}{Sm~C$^*$~}
\newcommand{\afe}{Sm~C$^*_{A}$~}

\newcommand{\fihigh}{Sm~C$^*_{FI2}$~}
\newcommand{\alfa}{Sm~C$^*_{\alpha}$~}

\newcommand{\bta}{Sm~C$^*_{\beta}$~}
\input epsf

\begin{document}
\draft

\twocolumn[\hsize\textwidth\columnwidth\hsize\csname@twocolumnfalse\endcsname
\title{General phase diagram for antiferroelectric liquid crystals
in dependence on enantiomeric excess}
\author{$^{1,2}$E. Gorecka, $^1$D. Pociecha, $^3$M. \v Cepi\v c,$^3$B. \v Zek\v s,
 $^4$R. Dabrowski}
\address{
$^1$Chemistry Department, Warsaw University, Al. Zwirki i Wigury
101, 02-089 Warsaw, Poland;\\
$^2$Department of Microelectronics and Nanoscience, Chalmers University of Technology,
S-41296 G\"{o}teborg, Sweden;\\
$^3$J. Stefan Institute, Jamova
39,1000 Ljubljana, Slovenia;\\
$^4$Institute of Chemistry,
Military University of Technology, 00-908 Warsaw, Poland. }
\date{\today}
\maketitle
\begin{abstract}
The phase diagram of the prototype antiferroelectric liquid
crystal MHPOBC in dependence of enantiomeric excess was measured.
It was shown that the \bta phase in very pure samples is the
\fihigh phase with a four layer structure, and only after small
racemization it transforms into the ferroelectric \fe phase. The
phase diagram was theoretically explained by taking into account
longer range bilinear and short range biquadratic interlayer
interactions, that lead to the distorted clock structures and
first order transitions between them.
\end{abstract}
] \narrowtext

In some systems chiral properties can be transferred from a
molecular to a macroscopic level. The best examples of such
systems are liquid crystals which form chiral phases (cholesteric,
blue, twist grain boundary or polar smectic phases), if built of
chiral molecules. The variation of the enantiomeric excess
influences macroscopic properties of these phases like the period
of the modulation and the phase transition temperature or it can
even change the phase sequence. The striking example of the last
phenomenon are antiferroelectric liquid crystals where some phases
disappear with racemization\cite{fukuda}. For high optical purity
MHPOBC compound, the prototype antiferroelectric material, the
sequence of tilted phases Sm~C$_{\alpha }^{\ast }$~
$\leftrightarrow $ Sm~C$_{\beta }^{\ast }$~
$\leftrightarrow $ Sm~C$_{\gamma }^{\ast }$~ $\leftrightarrow $ Sm~C$%
_{A}^{\ast }$~ was reported with decreasing temperature\cite{first}. In the
partially racemized sample only two of these phases, Sm~C$_{\beta }^{\ast }$%
~ and Sm~C$_{A}^{\ast }$, remained. The Sm~C$_{A}^{\ast }$~ phase is the
phase with the antiferroelectric properties and antiparallel tilts in
neighboring layers, i.e., anticlinic phase. The Sm~C$_{\beta }^{\ast }$~
phase has been recognized as the ferroelectric synclinic Sm~C$^{\ast }$%
~phase. Almost ten years later structures of the other subphases have been
found by resonant x-ray scattering in a different compound (S-10OTBBB1M7)
\cite{mach}. In this compound the phase sequence with decreasing temperature
is: the Sm~C$_{\alpha }^{\ast }$~ phase with the periodical modulation of a
tilt direction and with 5- or more, in general incommensurate, number of
layers, the Sm~C$_{FI2}^{\ast }$~ phase with 4-layer modulation, the Sm~C$%
_{FI1}^{\ast }$~ phase with 3-layer modulation, and the
Sm~C$_{A}^{\ast }$~ phase with 2-layer structure. The
Sm~C$_{FI1}^{\ast }$~ phase in the 10OTBBB1M7 compound corresponds
to the Sm~C$_{\gamma }^{\ast }$~ phase in MHPOBC material. So far
theoretical considerations failed to account for the correct phase sequences
in dependence of enantiomeric excess as well as experimentally consistent
structures of some phases.

In this Letter we present the complete phase diagram for the
antiferroelectric system MHPOBC with respect to the enantiomeric
excess. We show that the synclinic Sm~C$^{\ast }$ ~ phase
transforms into the \fihigh phase upon increasing optical purity. The result
solves the long known controversy why the smectic Sm~C$_{\beta
}^{\ast }$~ phase in MHPOBC was first reported as a typical
ferroelectric phase \cite{first} and later also as the phase
having antiferroelectric properties \cite{jakli}. We
show here, that this system is a nice example of intermediate
phases, which appear between main phases because of the cancelation
of short range interactions and consecutive relevance of
small chiral and/or longer range interactions. We present a new
phenomenological model, which takes into account longer range
bilinear \cite{flexo} and short range biquadratic \cite{reentrant}
interlayer interactions and allows for correct enantiomeric
excess dependent phase sequence with first order transitions
between experimentally consistent structures.

Several S-enantiomer rich \cite{dabrowski} mixtures of MHPOBC
were studied by differential scanning calorimetry (DSC),
dielectric and optical methods. The DSC measurements were
performed using Perkin Elmer DSC-7 calorimeter in the cooling and
heating runs at scanning rates 0.2-1~K/min. In the
dielectric measurements the glass cells of various thickness, with 25~mm$%
^{2} $ ITO electrodes coated by polimide, were used. Dielectric spectroscopy
studies were performed with HP 4192A impedance analyzer. In each phase the
dielectric spectra were fitted with Cole-Cole equation. The
selective reflection measurements were performed in transmission mode at
normal incidence (Shimadzu 3101PC spectrofotometer) for one surface free
(made on the quartz plate) or film samples. These studies allowed to
determine the helical pitch in Sm~C$_{A}^{\ast }$~and Sm~C$^{\ast }$~phases.

In the Sm~C$_{FI1}^{\ast }$and Sm~C$_{FI2}^{\ast }$~phases the pitch was
estimated by the direct microscopic observations of periodicity of\ the line
defects in 40 micron homogeneously aligned cell. The thick (250 micron) film
samples were used for the optical rotatory power (ORP) measurements, in
which a standard setup was used where the analyzer is rotated against the polarizer
to obtain a minimum of the light (630 nm) transmission.

In the optically pure samples (enantiomeric excess $x=c_S-c_R = \pm 1$ where $c_S$
and $c_R$ are corresponding concentrations)
four tilted smectic subphases with
historical names Sm~C$_{\alpha }^{*}$, Sm~C$_{\beta }^{*}$, Sm~C$_{\gamma
}^{*}$, Sm~C$_{A}^{*}$ appear below the Sm~A phase (Fig.~\ref{fig1}). Two
phases which appear between the Sm~C$_{\alpha }^{*}$ and Sm~C$_{A}^{*}$~
phase, have rather long helical pitch. The periodicity of disclination lines
which could be observed in the thick planar cell, is about 2.5~$\mu $m in
higher temperature phase (denoted as SmC$_{\beta }^{*}$) and about 1.6~$\mu $%
m in the lower temperature phase (denoted as SmC$_{\gamma }^{*}$). No
selective reflection in a visible range, which is encountered in the lower
purity MHPOBC materials, could be detected at any temperature. Both
intermediate phases have a pronounced optical activity, +35$^{o}$/$\mu $m
and -10$^{o}$/$\mu $m in SmC$_{\gamma }^{*}$ and SmC$_{\beta }^{*}$,
respectively, that is only slightly temperature dependent in the each phase
temperature interval. Except for the temperature region of SmC$_{\beta }^{*},
$ the results of ORP measurements agree with those presented in \cite
{skarabot}. Dielectric spectroscopy measurements clearly exclude the
presence of the ferroelectric synclinic Sm~C$^{*}$~ phase (Fig.~\ref{fig2}%
a). In the Sm~A phase, single, high frequency relaxation process is observed
that softens, e.g. the mode frequency decreases and its amplitude increases,
when approaching the Sm~C$_{\alpha }^{*}$~ phase. In the Sm~C$_{\alpha }^{*}$%
~ phase a single mode at $\sim $~60~kHz was observed. In the Sm C$_{\beta
}^{*}$ phase weak ($\Delta \varepsilon \sim 4$), high frequency (350kHz)
mode was detected, typical for the phase with an antiferroelectric order.
Similarly as in the Sm~C$_{A}^{*}$ phase \cite{panarin} this mode is either
related to the distortion of the crystallographic unit cell or to the rotation
of the molecules around their main axes. Absence of the Goldstone mode
excludes that Sm C$_{\beta }^{*}$ phase is the ferroelectric Sm C$^{*}$
phase.

Dielectric permittivity increases upon entering the Sm~C$_{\gamma }^{\ast }$%
~ phase. Here the main contribution to the dielectric susceptibility comes
from the low frequency relaxation process at 1-2 kHz, sometimes called the
ferrielectric Goldstone mode \cite{cgamma}. The relaxation frequency of the
mode detected in the Sm~C$_{\gamma }^{\ast }$~ phase is not well defined.
The parameter $\alpha $ in the Cole-Cole formula, that characterizes the
distribution of the relaxation frequencies, is about 0.3. The presence of a
broad mode is inherent to the Sm~C$_{\gamma }^{\ast }$~ phase. The
dielectric permittivity again decreases in the Sm~C$_{A}^{\ast }$~ phase, in
this phase two weak modes ($\Delta \varepsilon <1)$ could be seen in MHz
frequency region. Above results show that the Sm~C$_{\gamma }^{\ast }$~phase
is the ferrielectric phase identical as the Sm~C$_{FI1}^{\ast }$~and the Sm~C%
$_{\beta }^{\ast }$ phase is the antiferroelectric phase, biaxial thus
optically active and distinctly different than the Sm~C$_{A}^{\ast }$ phase.
Thus we conclude that the Sm~C$_{\beta }^{\ast }$ phase has the distorted
clock four-layer structure of the Sm~C$_{FI2}^{\ast }$ phase.

In the samples with a slightly lower optical purity ($x\sim 0.97$)
five subphases could be identified below SmA phase from the DSC
thermograms (Fig.\ref{fig1}) and microscopic studies. An additional phase,
which appears between the Sm~C$_{\alpha }^{\ast }$~ and the Sm~C$_{FI2}^{\ast }$%
~ phases, gives selective reflection in the visible light range. Selective
reflection wavelength changes from 450~nm to 600~nm within the phase
temperature interval. Since the temperature range of the Sm~C$_{FI2}^{\ast }$%
~ phase in this mixture is rather narrow it is difficult to observe all
five phases in the dielectric measurements. Due to the small temperature
gradients that are hard to avoid in the sample with the big electrode area,
the Sm~C$_{FI2}^{\ast }$~ phase always coexists with the additional phase.
In this additional phase the dielectric permitivity is an order of magnitude
stronger than in all other phases (Fig.\ref{fig2}b) thus typical for the
synclinic ferroelectric Sm C$^{\ast}$ phase.

As the optical purity decreases further, the Sm~C$_{FI2}^{\ast }$ phase
disappears and in the excess range between 0.88-0.97 the
phase sequence Sm~C$_{\alpha }^{\ast }$~ $\leftrightarrow $ Sm~C$^{\ast }$~ $%
\leftrightarrow $ Sm~C$_{FI1}^{\ast }$~ $\leftrightarrow $ Sm~C$_{A}^{\ast }$%
~ is observed, that is the phase sequence incorrectly reported in literature
for the pure enantiomeric MHPOBC compound \cite{first}. In mixtures with $x<0.88$
also Sm C$_{FI1}$ is missing. As
the enantiomeric excess is further reduced ($x<0.5$), finally the Sm~C$%
_{\alpha }^{\ast }$ disappears, and only Sm~A, Sm~C$^{\ast }$~ and Sm~C$%
_{A}^{\ast }$~ phases are left. Based on above observations the phase
diagram in dependence on enantiomeric excess as shown in Fig.~\ref{fig3}
is proposed. We believe that this type of the phase diagram is general
for antiferroelectric liquid crystals.

To account for experimental observations theoretically, we
introduce a new phenomenological model with bilinear interlayer
interactions between the tilt vectors $\bbox{\xi}_j$ which are of
longer range \cite{flexo}, and with biquadratic quadrupolar NN
interactions \cite{reentrant}, which become more important at
larger tilts i.e. at
 lower temperatures. The interlayer
part of the free energy is
\begin{eqnarray}
G_{int}=\frac{1}{2}\sum_{j} &&\left( \sum_{i=1}^{4}\tilde{a}_{i}\;\left( \xi
_{j}\cdot \xi _{j+i}\right) +\sum_{i=1}^{3}\tilde{f}_{i}\;\left( \xi
_{j}\times \xi _{j+i}\right) \right. +  \nonumber \\
&&+\left. b_{Q}\;\left( \xi _{j}\cdot \xi _{j+1}\right) ^{2}\right) .
\label{free}
\end{eqnarray}
Achiral parameters $\tilde{a}_{i}$  and chiral parameters
$\tilde{f}_{i}$ depend on steric and on van der Waals interactions
to nearest neighboring layers (NN), on electrostatic interactions to NN and
next nearest neighboring layers (NNN), on intralayer chiral piezoelectric
and NN flexoelectric coupling. They are given by Eq.(6) in the Ref. \cite{flexo}. We
assume that for the racemic mixture parameter $\tilde{a}_{1}$ is
negative favoring synclinic tilts at higher temperatures and
becomes positive favoring anticlinic tilts at lower temperatures.
In chiral samples it gains additional positive contribution proportional to $%
x^{2}$, thus changes the sign at higher temperature. Parameter $\tilde{a}_{2}$
is positive and does not depend on the enantiomeric excess considerably. Quadrupolar NN
interactions $b_{Q}$ are excess independent. All
parameters are expressed in degree Kelvin \cite {flexo}.

Numerical analysis of the above model gives the phase diagrams
which are in qualitative agreement with the experimentally
obtained one Fig. \ref{fig3}. This detailed analysis will be
published elsewhere and here only the basic physical
understanding is given. In racemic mixtures the synclinic negative
$\tilde{a}_1$ term prevails over the anticlinic positive
$\tilde{a}_2$ term directly below the transition from the Sm~A to
the tilted phase and the \fe phase appears. Upon decreasing
temperature, $\tilde{a}_1$ increases and changes its sign at a
temperature where quadrupolar term is already significant. Since
this term favors both synclinic and anticlinic tilts in NN layers,
direct transition from the \fe phase to the \afe phases is
obtained. In chiral samples, the anticlinic part of the parameter
$\tilde{a}_{1}$ increases and $\tilde{a}_{1}$ becomes  by its
absolute value comparable to the parameter $\tilde{a}_{2}$
directly below the transition to the tilted phase. This results in
a formation of the Sm~C$_{\alpha }^{\ast }$~ phase with the short
pitch modulation which becomes stable over a wider temperature
range upon increasing enantiomeric excess. It evolves into the
Sm~C$^{\ast }$~ phase due to increasing quadrupolar interactions.
Due to increased enantiomeric excess also the temperature increases
where the $\tilde{a}_{1} $ changes sign. Therefore the
influence of quadrupolar interactions is weaker and cannot prevent
the existence of modulated phases. With purification also the
temperature range of modulated phases increases. In the
temperature region where $\tilde{a}_{1}\approx 0$ the structure
with two interchanging phase differences $\alpha$ and $\beta$ with
$\alpha +\beta
\approx \pi $, is formed. The quadrupolar term $b_{Q}$ favors values of $%
\alpha =0$ and $\beta =\pi $ but chiral NN interactions increase
the value of $\alpha $ and decrease the value of $\beta$. This
structure obtained by the minimization of the Eq.~(\ref{free}) is
consistent with a tentatively proposed structure for the
Sm~C$_{FI2}^{\ast }$~ phase \cite{akizuki,johnson}. The structure
has the biaxial unit cell, a negligible polarization and is
helicoidally modulated with the long optical pitch defined by $\delta$
(Fig.\ref{fig4}a). Decreasing the temperature further, the parameter
$\tilde{a}_{1}$ becomes positive and thus simultaneously with the
$\tilde{a}_{3}$ parameter encourages the structure where tilts in
third neighboring layers are synclinic. Due to the up down
symmetry, the phase difference $\alpha$ is followed by two equal
phase differences $\beta$ where $\alpha + 2\beta \approx 2\pi $.
The obtained unit cell is polar and biaxial. The structure is
helicoidally modulated with the long optical pitch defined by $\delta$
(Fig.~\ref{fig4}b). It can be recognized as the one proposed for
the Sm~C$_{FI1}^{\ast }$~ phase \cite{akizuki,johnson}. In
pure samples the temperature range of the \alfa phase and the
temperature range of the modulated phases are so wide that the
Sm~C$^{*}$~ phase cannot evolve and direct phase transition
between the Sm~C$_{\alpha }^{*}$~ phase and the Sm~C$_{FI2}^{*}$~
phase occurs as observed in MHPOBC.

To conclude, in this Letter we report experimental and theoretical studies
which show that in the optically pure antiferroelectric liquid crystal
MHPOBC, the Sm~C$_{\beta }^{\ast }$~ phase is not the ferroelectric Sm~C$%
^{\ast }$ phase but the Sm~C$_{FI2}^{\ast }$ phase with four-layer
unit cell and antiferroelectric properties. The ferroelectric
Sm~C$^{\ast }$~ phase appears only after a slight racemization.
For antiferroelectric liquid crystals the general phase diagram
with respect to enantiomeric excess is proposed. We also proposed
the new free energy for polar smectics which includes achiral and
chiral bilinear interlayer interactions to more distant layers and
quadrupolar NN interactions. This free energy allows for the
correct enantiomeric excess dependent phase sequence as well as
the experimentally consistent structures of all phases and first
order transitions between them.

Acknowledgment: This work was supported by Polish-Slovenian
exchange program. The financial support for E.G. from $TFR$
Swedish Research Council is acknowledged. E.G. is gratefull to
Prof. S. Lagerwall for his hospitality during her stay at
the Chalmers University.

\begin{figure}[t]
\begin{center}
\epsfxsize=85mm \epsfbox{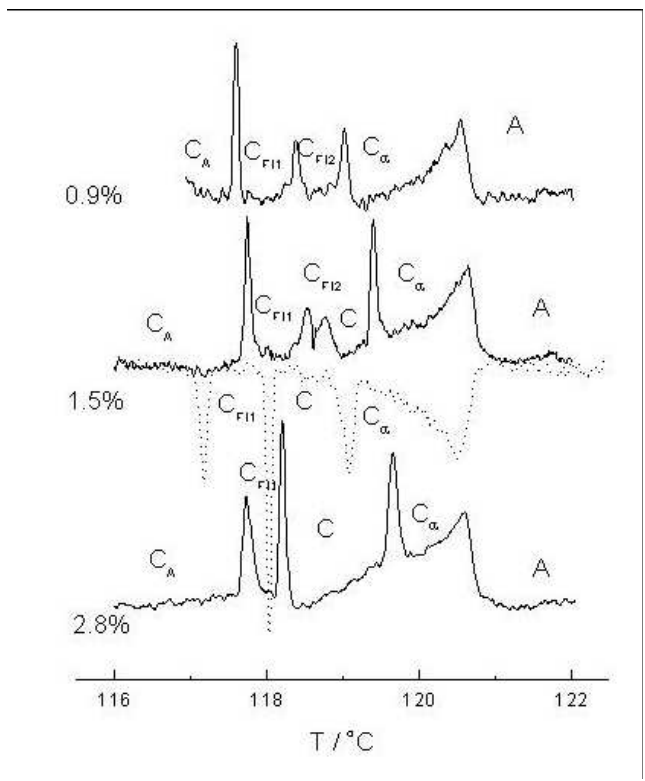}
\end{center}
\caption{DSC scans for MHPOBC mixtures with various percentage of
R enantiomer measured in heating scans with 1~K/min rate. For
concentration 1.5\% the cooling scan is also shown.} \label{fig1}
\end{figure}

\begin{figure}[t]
\begin{center}
\epsfxsize=85mm \epsfbox{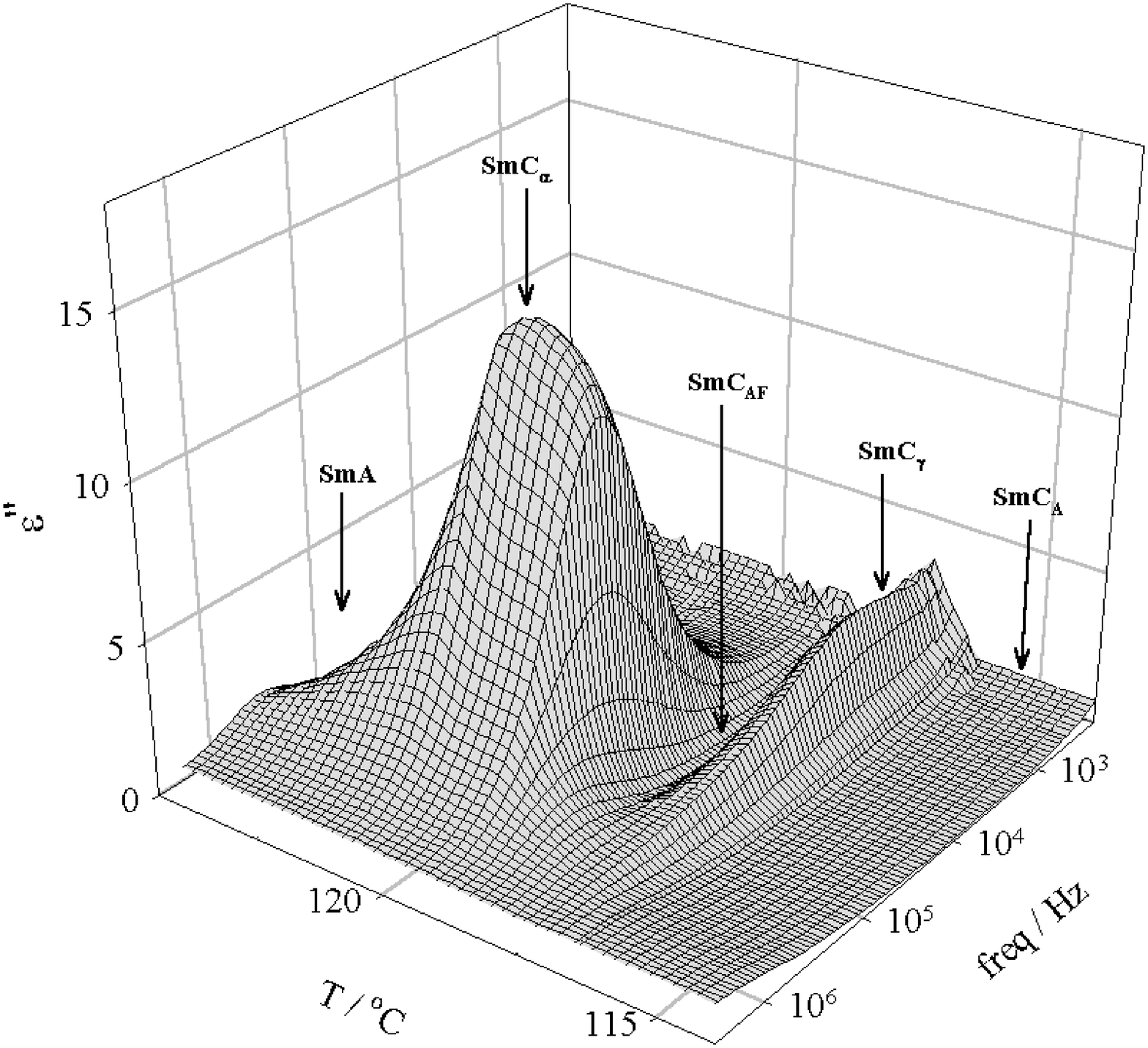}
\end{center}
\begin{center}
\epsfxsize=85mm \epsfbox{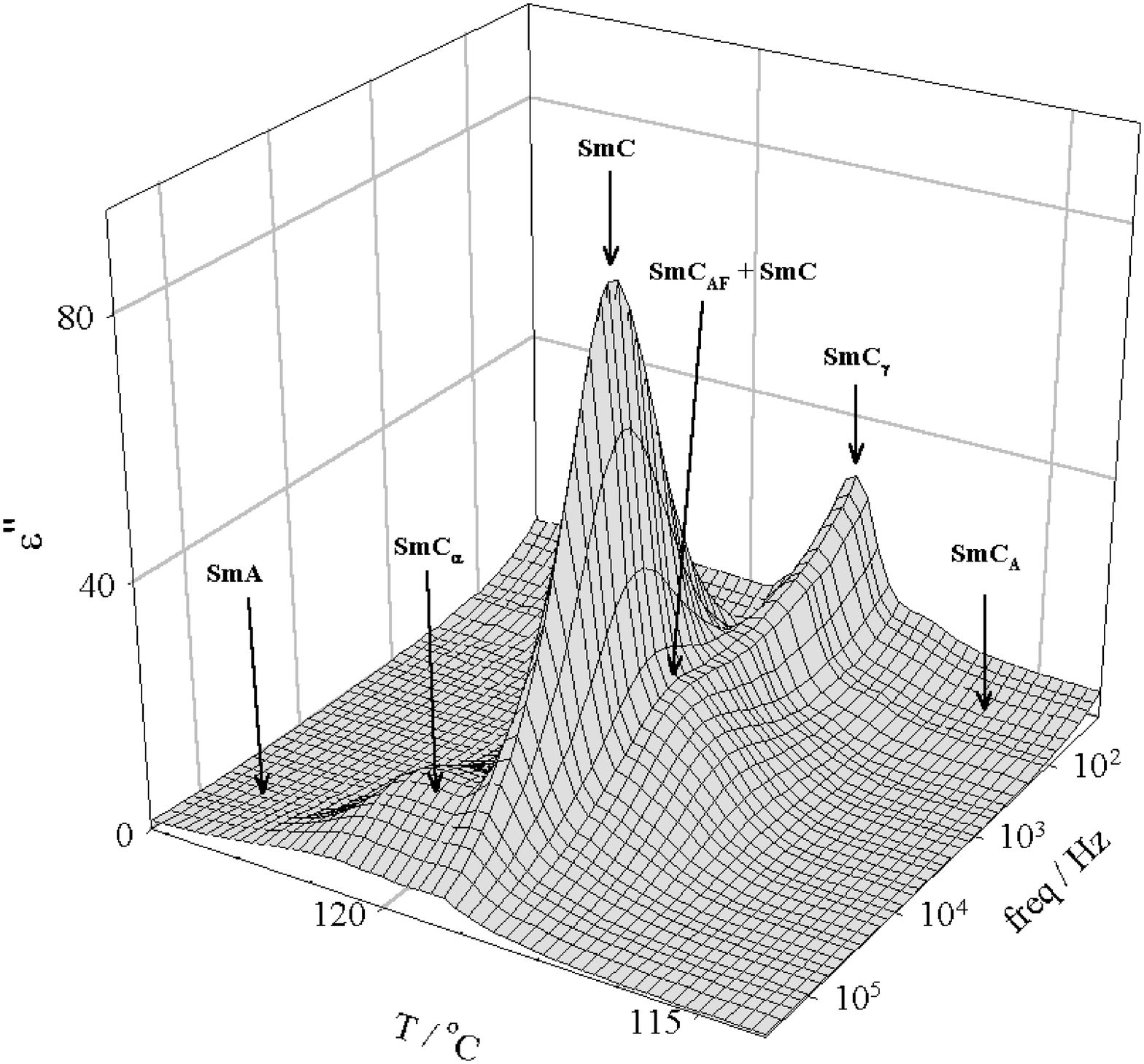}
\end{center}
\caption{Dielectric losses vs. temperature and frequency in
S-MHPOBC enantiomer (a) and and its mixture with 1,5\% of
R-enantiomer (b).} \label{fig2}
\end{figure}

\begin{figure}[t]
\begin{center}
\epsfxsize=85mm \epsfbox{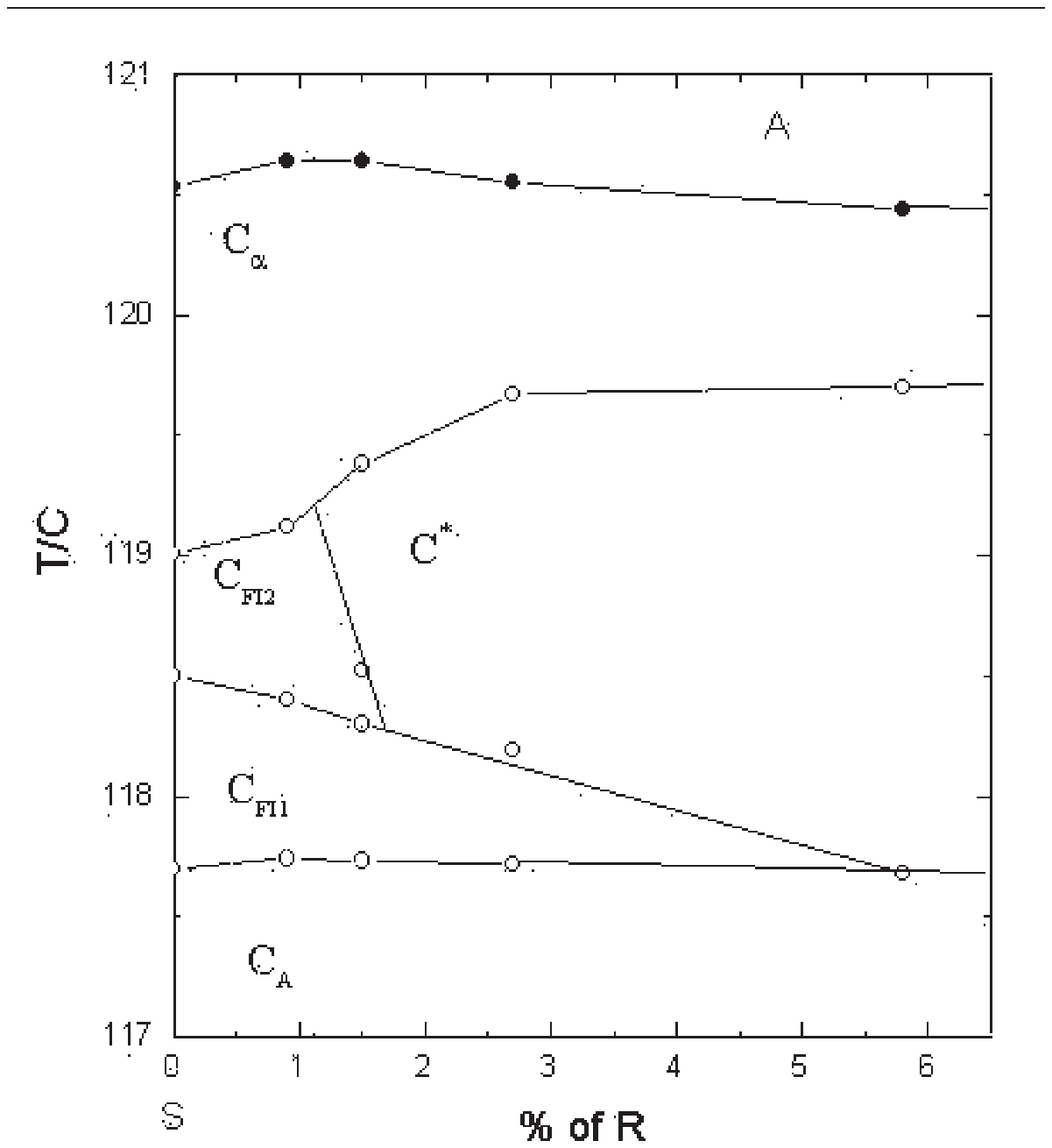}
\end{center}
\caption{Phase diagram for MHPOBC in dependence of optical
purity.} \label{fig3}
\end{figure}
\newpage
\begin{figure}[t]
\begin{center}
\epsfxsize=85mm \epsfbox{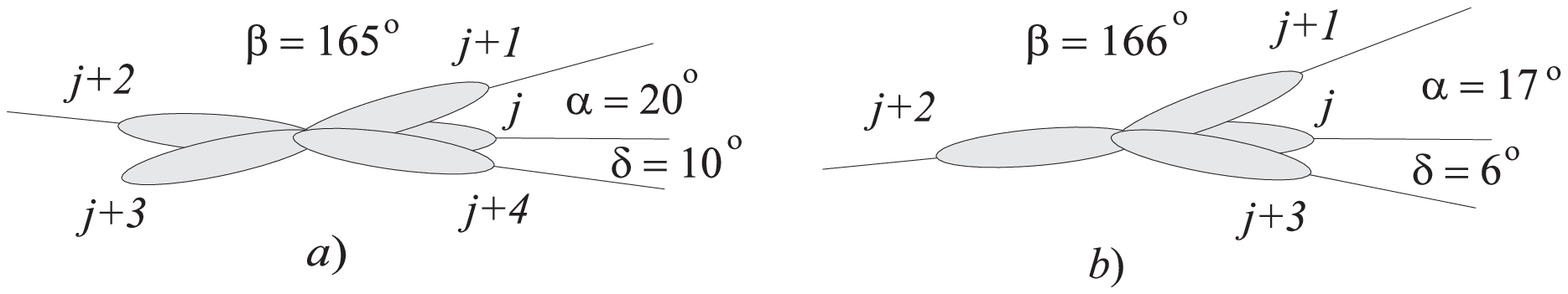}
\end{center}
\caption{a) The Sm~C$_{FI2}^{*}$~ phase and b) the
Sm~C$_{FI1}$~phase obtained for the optically pure material $x=1$
and the set of parameters expressed in K (Eq.(6), Ref.[5]): $a_1=
(-4.1 + 90 \; \theta^2)$, $c_p =0.6$, $\mu=0.77$, $b_0 = 2$, $b_1
= b_0/5$, $b_2 = b_0/100$, $f_1 = 0$ and for $b_Q = -17.7$.
Corresponding set of effective parameters expressed in mK is:
$\tilde{a}_2 = 73.5$, $\tilde{a}_3 = -7.41$, $\tilde{a}_4 = 0.65$,
$\tilde{f}_1 = -231$, $\tilde{f}_2 = 23.1$ and $\tilde{f}_3 =
2.02$. The tilt angle is taken from Ref.[8] for  (a) $\theta =
12.5^\circ$ and therefore $\tilde{a}_1=137$~mK and for (b) $\theta
= 13.2^\circ$ and $\tilde{a}_1=-69.6$~mK.}
\label{fig4}
\end{figure}


\begin{references}
\bibitem{fukuda}  A.Fukuda, Y.Takanishi, T.Isozaki, K.Ishikawa and
H.Takezoe, J.Mat.Chem. {\bf 4}, 997 (1994).

\bibitem{first}  A.D.L.Chandani, E.Gorecka, Y.Ouchi, H.Takezoe and A.Fukuda
Jap.J.App.Phys. {\bf 28}, L1265, (1989); E.Gorecka, A.D.L.Chandani, Y.Ouchi,
H.Takezoe and A.Fukuda, Jap.J.Appl.Phys. {\bf 29}, 131 (1990).

\bibitem{mach}  P.Mach, R.Pindak, A.-M.Levelut, P.Barois, H.T.Nguyen,
C.C.Huang, and L.Furenlid, Phys. Rev. Lett. {\bf 81}, 1015 (1998). P.Mach,
R.Pindak, A.-M.Levelut, P.Barois, H. T.Nguyen, H.Baltes, M.Hird, K.Toyne,
A.Seed, J.W.Goodby, C.C.Huang, and L.Furenlid, Phys.Rev.E {\bf 60},6793,
(1999).

\bibitem{jakli}  J.F.Li, E.A.Shack, Y.K.Yu, X.Y.Wang, C.Rosenblatt, M.E.
Neubert, S.S.Keats and H.Gleeson Jap. J.Appl.Phys. 2 {\bf 35}, L1608 (1996);
T.Sako,Y.Kimura, R.Hoyakawa, N.Okabe, Y.Suzuki, Jap.J.Appl.Phys. {\bf 235},
L114 (1996); A. Jakli J.Appl.Phys. {\bf 85}, 1101 (1999).

\bibitem{flexo}  M.\v {C}epi\v {c}, B.\v {Z}ek\v {s}, Phys.Rev.Lett 87,
085501 (2001).

\bibitem{reentrant}  D.Pociecha, E.Gorecka, M.\v {C}epi\v {c}, N.Vaupoti\v {c%
}, B.\v {Z}ek\v {s}, D.Kardas, and J.Mieczkowski, Phys.Rev.Lett. {\bf 86},
3048 (2001).

\bibitem{dabrowski}  R- or S- octanol-2 with 99.5\% enantiomeric excess was
used for the synthesis of the MHPOBC compound as described by W.J. Drzewinski Phd
thesis, (2000).

\bibitem{skarabot}  M.\v {S}karabot, M.\v {C}epi\v {c}, B.\v {Z}ek\v {s},
R.Blinc, G.Heppke, A.V.Kityk, I.Mu\v {s}evi\v {c}, Phys.Rev.E{\bf 58}, 575
(1998)

\bibitem{panarin}  Yu.P.Panarin, O.Kalinovskaya, J. K.Vij, J.W.Goodby
Phys.Rev.E{\bf 55}, 4345 (1997).

\bibitem{cgamma}  S.Merino, M.R.de la Fuente, Y.Gonz\'{a}lez, M.A.P\'{e}rez
Jubindo,~B.Ros, J.A.Pu\'{e}rtolas, Phys.Rev.E{\bf 54}, 5169 (1996); M.\v {C}%
epi\v {c}, G.Heppke, J.M.Hollidt, D.L\"{o}tzsch, B.\v {Z}ek\v {s},
Ferroelectrics {\bf 147}, 43 (1993); M.Glogarova, H.Sverenyuk, H.T.Nguyen,
C.Destrade, Ferroelectrics {\bf 147}, 159 (1993).

\bibitem{akizuki}  T.Akizuki, K.Miyachi, Y.Takanishi, K.Ishikawa, H.Takezoe
and A.Fukuda, Jpn.J.Appl.Phys. {\bf 38}, 4832(1999).

\bibitem{johnson}  P.M.Johnson, D.A.Olson, S.Pankratz, T.Nguyen, J.Goodby,
M.Hird and C.C.Huang, Phys.Rev.Lett.{\bf 84}, 4870 (2000).
\end{references}
\end{document}